\documentclass[twocolumn]{aastex61}
\shorttitle{GTC/\-Ca\-na\-ri\-Cam images of IC 418}
\shortauthors{D\'{\i}az-Luis et al.}

\begin{document}


\title{GTC/\-Ca\-na\-ri\-Cam MID-IR IMAGING OF THE FULLERENE-RICH PLANETARY NEBULA IC 418:
searching for the spatial distribution of fullerene-like molecules}




\author{J. J. D\'{\i}az-Luis}
\affil{Instituto de Astrof\'{\i}sica de Canarias, C/ Via L\'actea s/n, E-38205 La Laguna, Spain}
\affil{Departamento de Astrof\'{\i}sica, Universidad de La Laguna (ULL), E-38206 La Laguna, Spain}
\affil{Observatorio Astron\'omico Nacional (IGN), Alfonso XII, 3 y 5, 28014 Madrid, Spain}

\author{D. A. Garc\'{\i}a-Hern\'andez}
\affil{Instituto de Astrof\'{\i}sica de Canarias, C/ Via L\'actea s/n, E-38205 La Laguna, Spain}
\affil{Departamento de Astrof\'{\i}sica, Universidad de La Laguna (ULL), E-38206 La Laguna, Spain}

\author{A. Manchado}
\affil{Instituto de Astrof\'{\i}sica de Canarias, C/ Via L\'actea s/n, E-38205 La Laguna, Spain}
\affil{Departamento de Astrof\'{\i}sica, Universidad de La Laguna (ULL), E-38206 La Laguna, Spain}
\affil{Consejo Superior de Investigaciones Cient\'{\i}ficas, Madrid, Spain}

\author{P. Garc\'{\i}a-Lario}
\affiliation{Herschel Science Centre, European Space Astronomy Centre, European Space Agency,
Postbus 78, E$-$28691 Villanueva de la Ca\~nada, Madrid, Spain}

\author{E. Villaver}
\affiliation{Department of Theoretical Physics, Universidad Aut\'onoma de Madrid,
Cantoblanco, E$-$28049 Madrid, Spain}

\author{G. Garc\'{\i}a-Segura}
\affiliation{Instituto de Astronom\'ia, Universidad Nacional Aut\'onoma de M\'exico,
Km. 103 Carr. Tijuana-Ensenada, 22860, Ensenada, B. C., M\'exico}

\correspondingauthor{J. J. D\'{\i}az-Luis}
\email{jjairo@oan.es, agarcia@iac.es, amt@iac.es, Pedro.Garcia.Lario@esa.int, eva.villaver@uam.es, ggs@astro.unam.mx}


\begin{abstract}
We present seeing-limited narrow-band mid-IR GTC/\-Ca\-na\-ri\-Cam images of the
spatially extended fullerene-containing planetary nebula (PN) IC 418. The
narrow-band images cover the C$_{60}$ fullerene band at 17.4 $\mu$m, the
polycyclic aromatic hydrocarbon like (PAH-like) feature at 11.3 $\mu$m,
the broad 9-13 $\mu$m feature, and their adjacent continua at 9.8 and 20.5
$\mu$m. We study the relative spatial distribution of these complex
species, all detected in the {\it Spitzer} and {\it Infrared Space Observatory
(ISO)} spectra of IC 418, with the aim of getting observational constraints to
the formation process of fullerenes in H-rich circumstellar environments. A
similar ring-like extended structure is seen in all narrow-band filters, except
in the dust continuum emission at 9.8 $\mu$m, which peaks closer to the central
star. The continuum-subtracted images display a clear ring-like extended
structure for the carrier of the broad 9-13 $\mu$m emission, while the spatial
distribution of the (PAH-like) 11.3 $\mu$m emission is not so well
defined. Interestingly, a residual C$_{60}$ 17.4 $\mu$m emission (at about
4-$\sigma$ from the sky background) is seen when subtracting the dust continuum
emission at 20.5 $\mu$m. This residual C$_{60}$ emission, if real, might
have several interpretations; the most exciting being perhaps that other
fullerene-based species like hydrogenated fullerenes with very low
H-content may contribute to the observed 17.4 $\mu$m emission. We conclude
that higher sensitivity mid-IR images and spatially resolved spectroscopic
observations (especially in the Q-band) are necessary to get some clues
about fullerene formation in PNe.
\end{abstract}



\section{Introduction} \label{sec:int}

{\it Spitzer} observations have provided the first evidence for the presence of
C$_{60}$ and C$_{70}$ fullerenes in Planetary Nebulae (PNe; Cami et al. 2010;
Garc\'{\i}a-Hern\'andez et al. 2010, 2011a), reflection nebulae (Sellgren et al.
2010), the two least H-deficient R Coronae Borealis stars
(Garc\'{\i}a-Hern\'andez et al. 2011b) and a proto-PN (Zhang \& Kwok 2011). In
particular, the detection of C$_{60}$ fullerenes together with polycyclic
aromatic hydrocarbon like (PAH-like) features in PNe with normal
H-abundances indicate that formation of large fullerenes such as C$_{60}$ is
favored in the presence of H contrary to theoretical and experimental
expectations (Garc\'{\i}a-Hern\'andez et al. 2010). The detection of fullerenes
around old stars together with the recent identification of C$_{60}$$^{+}$ as a
diffuse interstellar band (DIB) carrier (the only one known to date; Campbell et
al. 2015; Walker et al. 2015) indicate that fullerenic molecular nanostructures
are ubiquitous in the interstellar medium (ISM), and could play a fundamental
role in many aspects of circumstellar/interstellar Chemistry and Physics.

The formation process of fullerenes in space is still uncertain and several
mechanisms have been proposed in the literature as the most suitable ones: i)
the formation in H-poor environments (Goeres \& Sedlmayr 1992); ii) high- and
low-temperature formation in C-rich environments (J$\ddot{a}$ger et al.
2009; Fulvio et al. 2017); iii) photochemical processing of hydrogenated
amorphous carbon grains (HACs; Garc\'{\i}a-Hern\'andez et al. 2010); and iv)
photochemical processing of large PAHs (Bern\'e \& Tielens 2012). Hydrogenated
species cannot be produced by the first two mechanisms, which is very difficult
to reconcile with the astronomical evidence. The two last formation scenarios,
involving the photochemical processing of HACs (or similar materials with a
mixed aromatic/aliphatic composition; e.g., Garc\'{\i}a-Hern\'andez et al. 2012)
and large PAHs, are based on top-down chemical models toward the most stable
C$_{60}$ and C$_{70}$ fullerenes.

C$_{60}$, in the Bern\'e \& Tielens (2012) model, is formed top-down from the
transformation of large PAHs into graphene under UV radiation from massive
stars. Potential problems of the large PAHs scenario above are that: i) yet
to date, no specific gas-phase PAH has been identified in space (see e.g.,
Tielens 2011 for a review); and ii) PAHs cannot explain the origin of the band
variations depending on the astrophysical source (e.g., Zhang \& Kwok 2015;
\'Alvaro Galu\'e \& D\'{\i}az Leines 2017). Interestingly, Zhen et al. (2014)
reported the first experimental evidence that PAHs with 60 C atoms
photo-isomerize to C$_{60}$, while Bern\'e et al. (2015) presented a detailed
photochemical modelling for a single PAH molecule (C$_{66}$H$_{20}$) that may
explain the the top-down formation of fullerenes in the ISM via UV processing of
large PAH molecules.

Contrary to the large PAHs fullerene formation scenario, the presence of
HAC-like materials in space has been firmly established; e.g., from the
detection of the sp$^{3}$ CH$_{3}$ and CH$_{2}$ stretching features contributing
to the 3.4 $\mu$m absorption detected in very different astronomical
environments (e.g., Alata et al. 2014 and references therein) or the aliphatic
discrete emission features (e.g., at 3.4, 3.5, 6.0, 6.9, and 7.3 $\mu$m) and
broad emission plateaus (e.g., at 6-9 and 9-13 $\mu$m) (e.g., Kwok \& Zhang 2011
and references therein) associated to HAC-like materials. Fullerenes and
PAH-like species in H-containing circumstellar environments may be formed by
the decomposition (e.g., by the UV radiation from the central star) of a
carbonaceous compound with a mixture of aromatic and aliphatic structures
(HAC-like; Garc\'{\i}a-Hern\'andez et al. 2010, 2011a,b, 2012; Micelotta et al.
2012), which should be a major constituent in the circumstellar envelope of
C-rich evolved stars. Potential problems to this alternative fullerene
formation scenario (the HAC's scenario) are: i) no specific HAC-like particle
has been identified in space (e.g., Kwok \& Zhang 2011); and ii) the low content
of aliphatic C relative to aromatic C apparently estimated from astronomical
observations (e.g., Li \& Draine 2012). However, the HAC's fullerene formation
scenario is suggested by the coexistence of a complex mix of aliphatic and
aromatic species such as HACs, PAH clusters, fullerenes, and small
dehydrogenated carbon clusters (possibly planar C$_{24}$ or a small fragment of
a graphene sheet) in PNe of the Magellanic Clouds and our own Galaxy
(Garc\'{\i}a-Hern\'andez et al. 2011a, 2012). The coexistence of these molecular
species in PNe with fullerenes is supported by the laboratory experiments
carried out by Scott et al. (1997), which showed that the decomposition of HACs
is sequential; small dehydrogenated PAH molecules are released first, followed
by fullerenes and large PAH clusters. This complex top-down formation process is
also supported by the unique spectral variations seen in the IR spectrum of the
fullerene-containing RCB star V854 Cen, which indicate that HACs have evolved
into complex species such as PAH-like species and fullerene molecules in a
timescale of only a decade (Garc\'{\i}a-Hern\'andez et al. 2011b). 

In short, both top-down fullerene formation models are quite similar, the
main difference being the starting ingredient (HAC-like materials vs. PAHs).
Perhaps both models may work in space depending on the astrophysical environment
or they could be somekind related; e.g., it could be possible that
HAC$->$PAH$->$Graphene$->$C$_{60}$ under UV photon irradiation. Interestingly,
very recently \'Alvaro Galu\'e \& D\'{\i}az Leines (2017) show that the inclusion
of nonplanar structural defects in aromatic core molecular
structures\footnote{Note that defective out-of plane aromatic-like structures
are 3D species, not falling under the standard (2D) PAHs chemical definition
(see \'Alvaro Galu\'e \& D\'{\i}az Leines 2017).} can explain for the first time
the 6$-$9 $\mu$m emission spectral band patterns (i.e., the 6.2, 7.7, and 8.6
$\mu$m features usually attributed to PAHs) depending on astrophysical source.
They propose that the global regularity of the 6$-$9 $\mu$m IR emission bands is
consistent with an emission from the delocalized sp$^{2}$ phase (as $\pi$
domains) confined in mixed aromatic-aliphatic disordered carbon aggregates
(e.g., similar to coal, petroleum fractions, HAC, etc.). The photomechical
processing of HAC-like and PAH-like grains could thus lead to a defective aromatization,
being implicated in the top-down formation of fullerenes.

What is the prevalent formation process of fullerenes in
circumstellar/interstellar environments? The determination of the relative
spatial distribution of fullerenes, PAH-like species, and the yet
unidentified 9-13 $\mu$m feature carrier may represent the ideal test for
answering this question and for understanding the chemistry of large organic
molecules in circumstellar/interstellar environments. {\it Spitzer} observations
of fullerene-containing PNe only offer a marginal amount of spatial information
at a resolution of 3.6" (spatial information is only possible for
low-resolution, R$\sim$100, spectroscopy) which does not allow to determine the
exact relative spatial distribution of these species. However, seeing-limited
Gran Telescopio de Canarias (GTC) observations can improve the spatial
resolution by a factor of about 10 with respect to {\it Spitzer}. 

In this paper, we present GTC/\-Ca\-na\-ri\-Cam mid-IR imaging of the extended
fullerene-containing PN IC 418. An overview of the observations (and data
reduction) and of the dust continuum-subtraction techniques applied to our
images are presented in Sections 2 and 3, respectively. In Section 4 we study
the relative spatial distribution of the fullerene, PAH-like, and 9-13
$\mu$m emissions. Section 5 discusses these results, and the main conclusions of
our work are given in Section 6.

\section{Observations and data reduction} \label{sec:obs}

IC 418 is an elliptical PN, with a major axis of 14 arcsec and a minor axis of
11 arcsec in the optical (Ramos-Larios et al. 2012), which is surrounded by a
more extended low-level ionized halo, and, in turn, by a neutral envelope with
an angular size of about 2 arcmin (Taylor \& Pottasch 1987; Taylor et al. 1989).
This PN displays fullerene and PAH-like (e.g., at 11.3 $\mu$m) features,
together with a strong and broad 9-13 $\mu$m feature (see Fig. 1). Moreover, we
selected IC 418 (with an estimated distance of 1.26 kpc; Morisset \& Georgiev
2009) for the GTC/\-Ca\-na\-ri\-cam observations because it is the brightest and
most spatially extended source from the known list of Northern
fullerene-containing PNe (e.g., Garc\'{\i}a-Hern\'andez et al. 2012; Otsuka et
al. 2014). 

\begin{figure*}[ht]
%Figure 1
\centering
\includegraphics[angle=0,scale=.75]{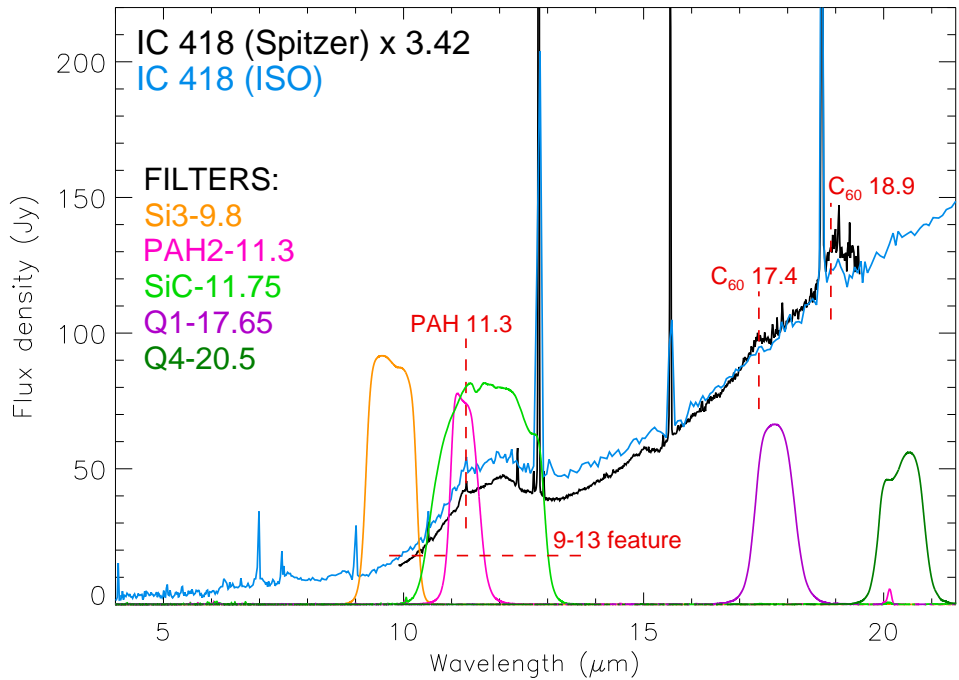}
\caption{{\it ISO} (in blue) and {\it Spitzer} (in black) IR spectra of the PN
IC 418 plotted together with the profiles of the five \-Ca\-na\-ri\-Cam filters used in
the observations. The positions of the C$_{60}$ bands at 17.4 and 18.9 $\mu$m,
and the PAH-like feature at 11.3 $\mu$m (red dashed vertical lines) are marked.
The broad 9-13 $\mu$m feature (red dashed horizontal line) is also marked.
\label{fig1}}
\end{figure*}

IC 418 was observed at the GTC with the mid-infrared camera \-Ca\-na\-ri\-Cam (Telesco et
al. 2003) in service mode in January and December 2013. \-Ca\-na\-ri\-Cam can perform
mid-infrared (7.5-25 $\mu$m) imaging using a set of narrow, medium and broad-band
filters. The seeing, in the V band, during the observations was 0.9" and 1.7" for the N
(10 $\mu$m) and Q (20 $\mu$m) images, respectively. This corresponds to a FWHM of 0".66
and 1".2 for the N and Q bands, respectively. Therefore the images are seeing limited.
These mid-IR images were obtained by using narrow-band filters centered around the
position of the features of our interest: PAH-like at 11.3 $\mu$m (PAH2, $\Delta$$\lambda$
$=$ 0.6 $\mu$m), 9-13 $\mu$m at 11.75 $\mu$m (SiC, $\Delta$$\lambda$ $=$ 2.5 $\mu$m), and
C$_{60}$ at 17.65 $\mu$m (Q1, $\Delta$$\lambda$ $=$ 0.9 $\mu$m), as well as their
adjacent continua at 9.8 $\mu$m (Si3, $\Delta$$\lambda$ $=$ 1.0 $\mu$m), and 20.5 $\mu$m
(Q4, $\Delta$$\lambda$ $=$ 1.0 $\mu$m). 

In order to map the inner regions ($<$20"), we have used the off-chip chopping and
nodding techniques, that consist of introducing small offsets in multiple images to
remove most of the background emission, the thermal emission from the telescope, and the
detector noise. Moreover, the off-chip mode means that the negative images (from the
off-source chop beam) are not seen in the detector image. The chop and nod position
angles were 75$^\circ$ and $-$105$^\circ$, respectively, with throws of 40 arcsec. We got
one final summed image of IC 418 in the Si3, PAH2, and SiC filters, while we averaged 2
and 3 summed images in the Q1 and Q4 filters, respectively, in order to get a final image
for each Q-band filter. 

Because we did not observe standard stars for flux calibration on the same nights as PN
IC 418, we have followed two different procedures for flux calibration: i) using the
observations of standard stars (HD 70272 and HD197989; available from the GTC data
archive) that were observed at observational conditions as similar as possible to those
of IC 418 but on different dates; and ii) using the {\it Infrared Space Observatory}
({\it ISO}) spectrum that covers the full nebula and assuming no significant IC 418 IR
spectral variability in the last $\sim$30 years (see below). In the first case, the final
images were processed using the pipeline RedCan (Gonz\'alez-Mart\'in et al. 2013), which
produces final flux-calibrated (in Jy) images from the GTC/\-Ca\-na\-ri\-Cam raw data.
Since the observations of the standard stars were made on different dates, in principle,
the flux calibration may be quite uncertain; although we get $\sim$10$-$30\% (depending
on the filter) relative flux differences between GTC/\-Ca\-na\-ri\-Cam and {\it ISO}. In
the second case, we first calculated the synthetic photometry (in Jy)  through all
\-Ca\-na\-ri\-Cam filters from the {\it ISO} spectrum using the SMART\footnote{The
Spectroscopic Modeling Analysis and Reduction Tool (SMART), is a software package to
reduce and analyze data from the Infrared Spectrograph (IRS) on the {\it Spitzer Space
Telescope}. Cornell maintains information on their website for SMART and updates it
regularly. See http://isc.astro.cornell.edu/smart.} software and we then calibrated our
mid-IR images in order to get a total summed flux (in Jy) that coincides with the
synthetic photometry. In Figure 2, we show the {\it ISO} spectrum of IC 418 in comparison
to the observed WISE (Cutri et al. 2013), AKARI (Ishihara et al. 2010), and IRAS (Helou
\& Walker 1988) photometry, as well as the corresponding synthetic photometry (again
obtained from the {\it ISO} spectrum using SMART) in each one of the WISE, AKARI, and
IRAS filters. There is no significant flux variation in the last $\sim$30 years; the only
exception is the WISE W4 flux density, which seems to deviate from the {\it ISO} spectrum
likely due to known saturation issues. Table 1 gives the filter names, total on-source
exposure times, signal-to-noise ratio (S/N), observational dates, integrated fluxes,
synthetic photometry, and the PN size in each mid-IR filter as derived from our images.

\begin{figure*}[ht]
%Figure 2
\centering
\includegraphics*[angle=0,scale=.75]{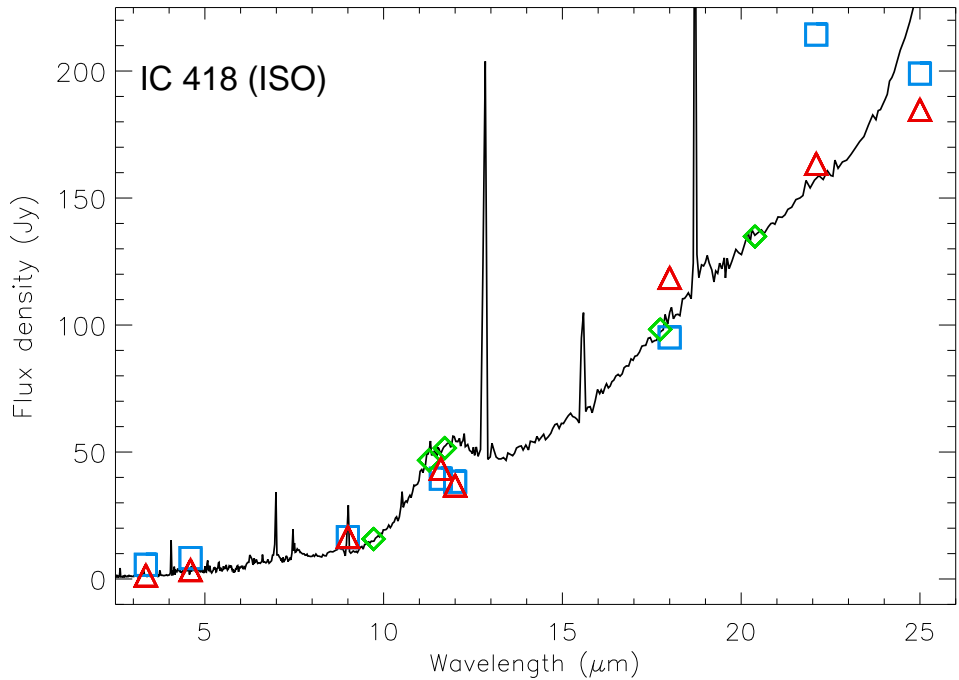}
\caption{{\it ISO} spectrum of the PN IC 418 in comparison to observed (squares)
and synthetic (triangles) {\it AKARI}, {\it WISE}, and {\it IRAS} photometry.
The GTC/\-Ca\-na\-ri\-Cam (diamonds) synthetic photometry is also displayed for
comparison.
\label{fig2}}
\end{figure*}

\begin{deluxetable*}{lccccccccccc}
\tablenum{1}
\tablecaption{Summary of observational data}
\tablehead{
\colhead{Filter\tablenotemark{a}} & \colhead{$\lambda$$_{c}$ ($\mu$m)} &
\colhead{Time\tablenotemark{b} (s)} & \colhead{S/N\tablenotemark{c}} &
\colhead{Date} & \colhead{Flux\tablenotemark{d} (Jy)} &
\colhead{SP\tablenotemark{e} (Jy)} & Size\tablenotemark{f} \\}
\startdata
 Si3 (cont\tablenotemark{g}) & 9.8 & 661  & 13 & 2013 Jan 20 & 4.52  & 15.76 & 13.6" x 15.5" \\
 PAH2 (PAH-like $+$ 9-13 $\mu$m $+$ cont\tablenotemark{g}) & 11.3 & 626  & 15 & 2013 Jan 20 & 36.76 & 46.78 & 14.3" x 17.4" \\
 SiC (PAH-like $+$ 9-13 $\mu$m $+$ cont\tablenotemark{g}) & 11.75 & 413  & 13 & 2013 Jan 2  & 42.39 & 51.56 & 13.8" x 16.0" \\
 Q1 (C$_{60}$ $+$ cont\tablenotemark{g}) & 17.65 & 1274 & 10 & 2013 Dec 29 & 93.93 & 98.28 & 14.2" x 16.0" \\
 Q4 (cont\tablenotemark{g}) & 20.5 & 1911 & 12 & 2013 Dec 29-30 & 112.69 & 134.91 & 14.6" x 17.1" \\
\enddata
\tablenotetext{a}{These are the names given to each \-Ca\-na\-ri\-Cam filter.}
\tablenotetext{b}{Total on-source exposure time.}
\tablenotetext{c}{S/N is an average of several measurements in each image.}
\tablenotetext{d}{Integrated fluxes when calibrating with the standard stars HD
70272 and HD197989 (see text for more details). The fluxes were measured
defining the PN size in each mid-IR filter at the 3$\sigma$ level (from the
mean sky background) using the Graphical Astronomy and Image Analysis Tool
(GAIA; http://star-www.dur.ac.uk/$\sim$pdraper/gaia/gaia.html). Flux errors are
estimated to be in the range 10-30\% (depending on the filter).}
\tablenotetext{e}{Synthetic photometry through all \-Ca\-na\-ri\-Cam filters from the
{\it ISO} spectrum using SMART (see text for details).}
\tablenotetext{f}{Sizes measured at the 3$\sigma$ level (from the mean sky
background) in each mid-IR filter with standard tasks in IRAF.}
\tablenotetext{g}{Dust continuum emission.}
\label{tab:observations}
\end{deluxetable*}

Figures 3 and 4 show the final flux-calibrated mid-IR GTC/\-Ca\-na\-ri\-Cam images of IC
418 using the synthetic photometry (case ii above). The PAH2 and SiC filters (with
different widths) cover the PAH-like feature at 11.3 $\mu$m, the broad 9$-$13 $\mu$m feature,
and dust continuum emission, while the Q1 filter covers the C$_{60}$ emission band at
17.4 $\mu$m and dust continuum emission. The Si3 and Q4 filters cover the dust continuum
emission at 9.8 and 20.5 $\mu$m, respectively (see Fig. 1). In short, a similar ring-like
extended structure is seen in all GTC/\-Ca\-na\-ri\-Cam filters, with the exception of
the Si3 filter that shows that the dust continuum emission at 9.8 $\mu$m peaks closer to
the central star. Note that we can exclude any significant contribution from the intense
emission line of Ne\,{\sc ii} at 12.81 $\mu$m to the SiC filter, as this line has only a
contribution of $\sim$8.2$\%$ to the synthetic photometry in that filter. Moreover, the
similar distribution of the PAH2 and SiC images suggests that we are seeing the emission,
but may not be contributing too much.

\begin{figure*}
%Figure 3
\centering
\includegraphics*[angle=0,scale=.58]{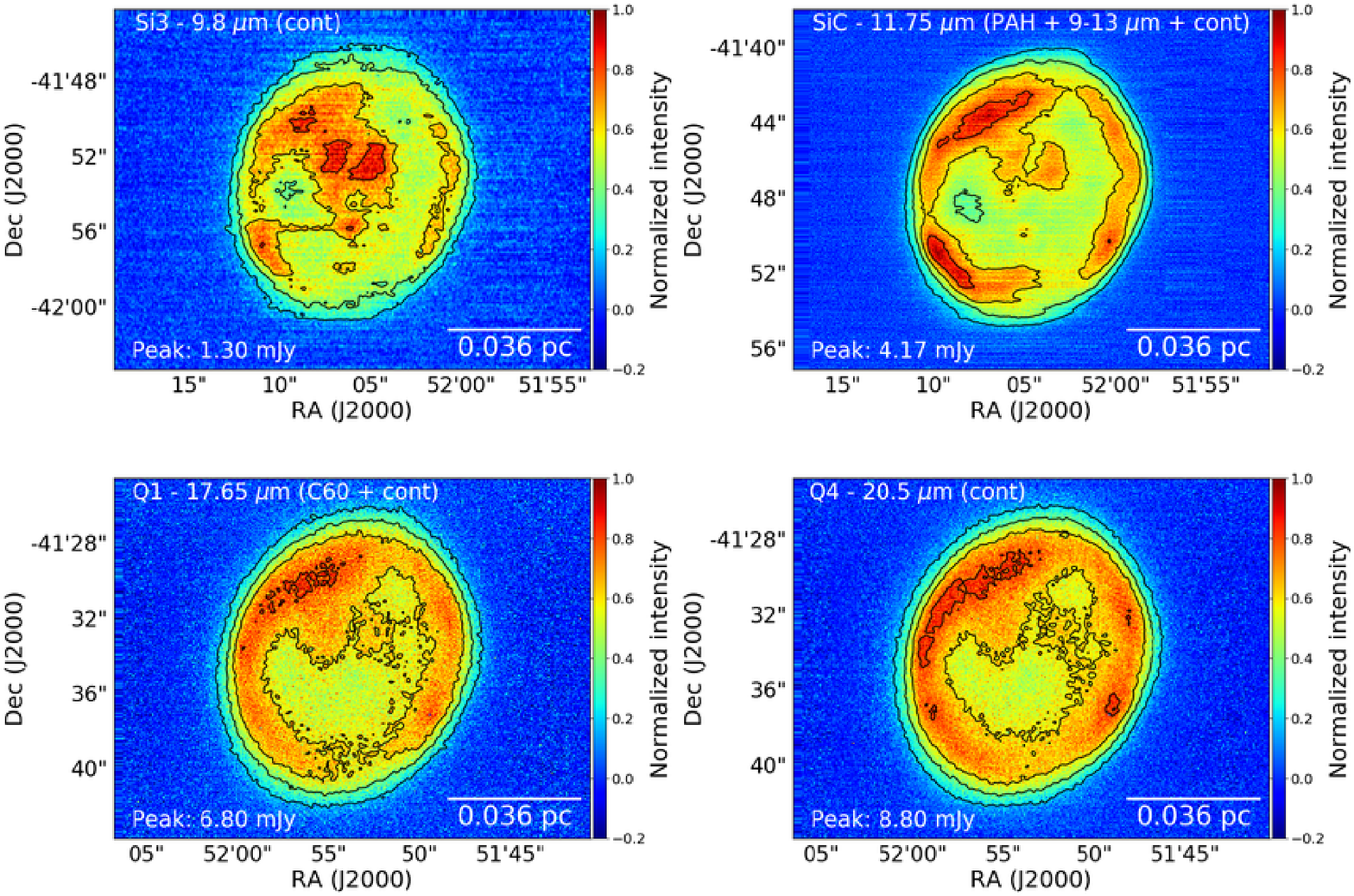}
\caption{Contour maps of the flux-calibrated mid-IR GTC/\-Ca\-na\-ri\-Cam images of the
C$_{60}$-PN IC 418 in the Si3 (continuum at 9.8 $\mu$m), SiC
(PAH-like $+$ 9-13 $\mu$m $+$ continuum at 11.75 $\mu$m), Q1 (C$_{60}$ $+$ continuum at
17.65 $\mu$m), and Q4 (continuum at 20.5 $\mu$m) filters. All contour maps have
been normalized to the peak flux in the image. North is up, east is left. The
bar in the lower right corner illustrates 0.036 pc at the estimated distance to IC 418
(1.26 kpc; Morisset \& Georgiev 2009). Contours range from 0.2 to 0.8 with 3
steps of 0.2 each. \label{fig3}}
\end{figure*}

\begin{figure}[h]
%Figure 4
\centering
\includegraphics*[angle=0,scale=.22]{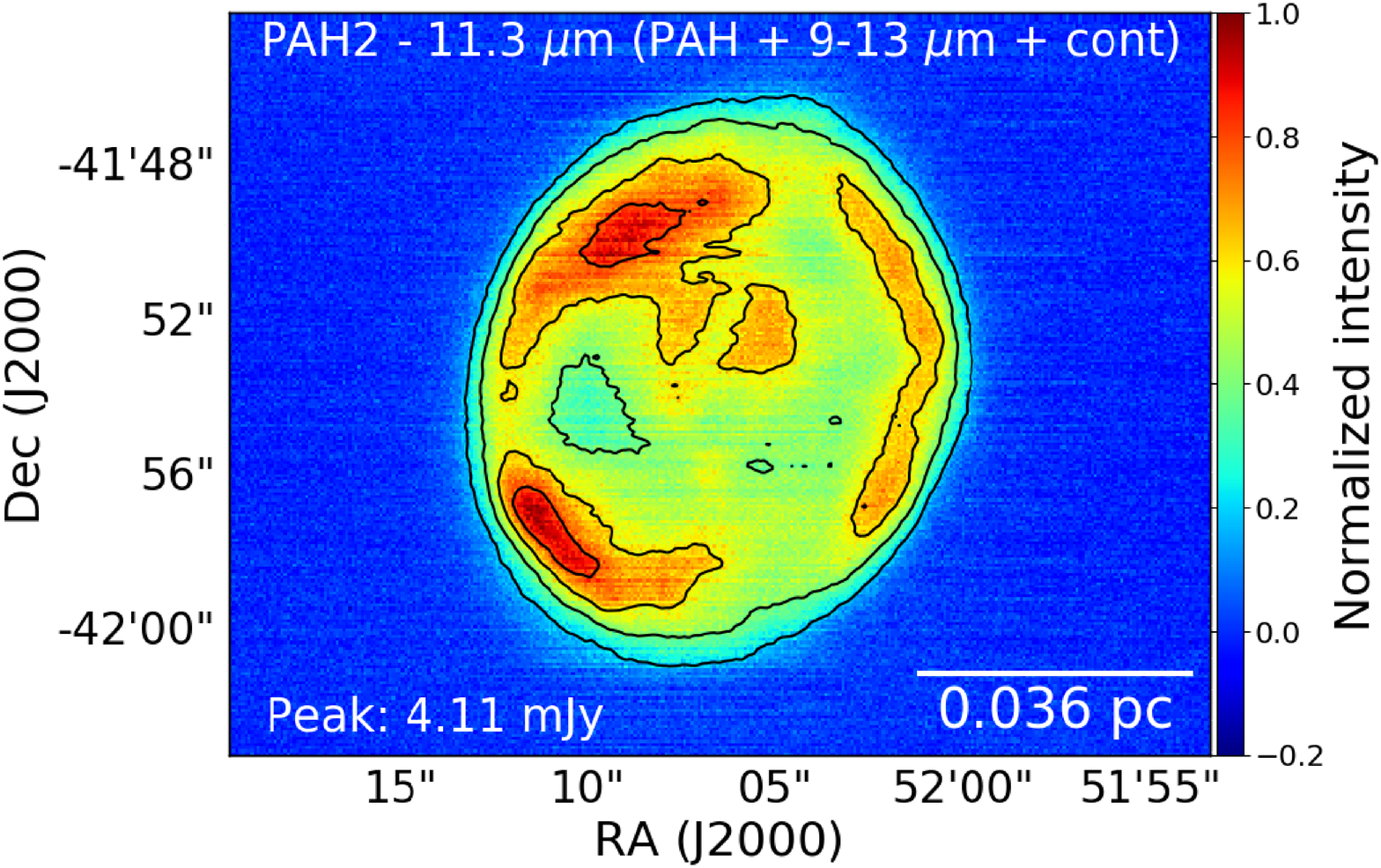}
\caption{Contour map of the flux-calibrated mid-IR GTC/\-Ca\-na\-ri\-Cam image of the
C$_{60}$-PN IC 418 in the PAH2 (PAH-like $+$ 9-13 $\mu$m $+$ continuum at 11.3 $\mu$m)
filter. The notation is the same as in Figure 3.
\label{fig4}}
\end{figure}

Finally, narrow-band Hubble Space Telescope (HST) WFPC2 images of IC 418 in the [N\,{\sc ii}]
$\lambda$6583 and [O\,{\sc iii}] $\lambda$5007 emission lines were retrieved from the
Mikulski Archive for Space Telescopes (MAST; Prop. ID: 8773) for comparison with
our mid-IR images. 

The 200-s [O\,{\sc iii}] image was obtained from one individual exposure acquired through
the F502N filter (central wavelength $\lambda$$_{c}$ = 5012 \AA\ and bandwidth
$\Delta$$\lambda$ = 27 \AA), whereas the 300-s [N\,{\sc ii}] image was obtained from one
individual exposure acquired using the F658N filter ($\lambda$$_{c}$ = 6591 \AA,
$\Delta$$\lambda$ = 29 \AA). Cosmic rays were removed using the Lacos$\_$im package of
van Dokkum (2001). 

\section{Dust continuum-subtraction} \label{sec:dust}

In order to study the relative spatial distribution of C$_{60}$ fullerenes,
PAH-like molecules, and the 9-13 $\mu$m feature carrier, we have subtracted the
dust continuum contribution by using the Si3 (9.8 $\mu$m) and Q4 (20.5 $\mu$m)
images.

\begin{figure*}
%Figure 5
\centering
\includegraphics*[angle=0,scale=.63]{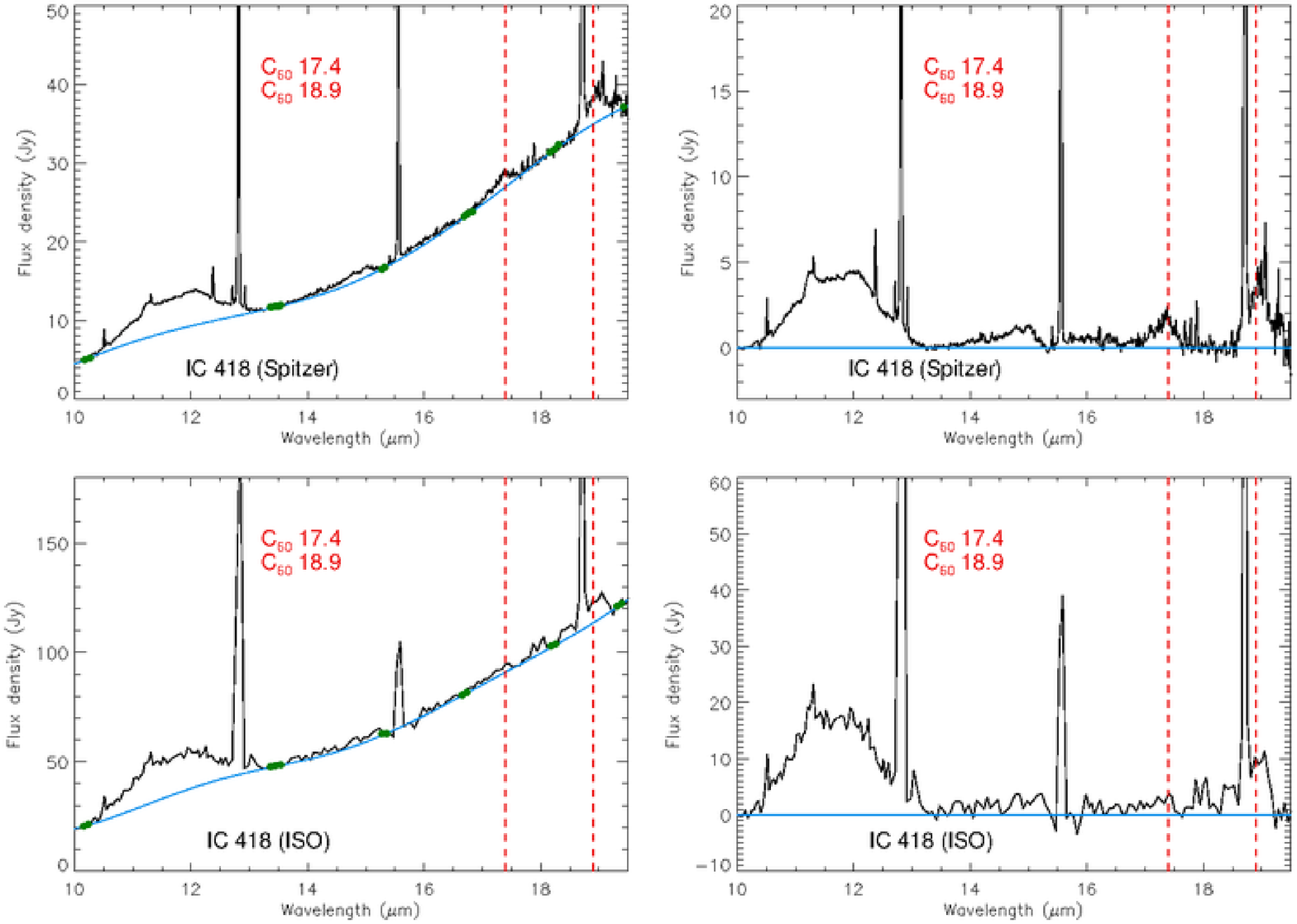}
\caption{Polynomial fits made to the dust continuum of the {\it Spitzer}/IRS
and {\it ISO} spectra (left panels) and the corresponding residual spectra
(right panels) for the fullerene PN IC 418. Both left panels display the
observed IR spectra (in black) together with a polynomial fit (in blue) to
continuum points free from any gas and dust feature (in green). The corresponding residual
or dust continuum subtracted spectra (in black) are shown in right panels. The
band positions of the C$_{60}$ features at 17.4 and 18.9 $\mu$m (red dashed
vertical lines) are indicated. \label{fig5}}
\end{figure*}

The continuum subtraction from our images is complicated because the central star of PN
IC 418 (nor any field star) is not detected in our mid-IR images. Since our
flux-calibrated images and the {\it Spitzer} and {\it ISO} spectra of IC 418 are in the
same flux units (Jy), we have then only taken into account the slope of these spectra;
i.e., the relative differences between the measured fluxes in the continuum at the
central wavelength in all filters are used to scale the continuum images and subtract the
dust continuum contribution\footnote{The relative flux differences are very similar in
the {\it Spitzer} and {\it ISO} spectra (see Fig. 1).}. In order to estimate the dust
continuum contribution for each filter, we have fitted a polynomial spline of order 4 to
the continuum (free from any gas/dust feature) in the {\it Spitzer} and {\it ISO} spectra
by using standard tasks in IRAF\footnote{Image Reduction and Analysis Facility (IRAF)
software is distributed by the National Optical Astronomy Observatories, which are
operated by the Association of Universities for Research in Astronomy, Inc., under
cooperative agreement with the National Science Foundation.} (see Fig. 5).

By assuming that the continuum flux ratios (or relative differences) between the
wavelengths of our interest are similar in both {\it Spitzer/ISO} spectra and
GTC/\-Ca\-na\-ri\-Cam images (see above about the non IR variability of IC 418
during the last $\sim$30 years; Fig. 2), we find that we have to scale the Q4
image by a factor of $\sim$0.70 for the case of the Q1$-$Q4
continuum-subtraction. The Si3 images have to be scaled by factors of $\sim$1.90
and 2.17 for the PAH2$-$Si3 and SiC$-$Si3 continuum-subtractions, respectively.
Moreover, in order to extract the PAH-like 11.3 $\mu$m emission from the
PAH2 image (PAH2$-$SiC), we have to scale the SiC image (assumed to be dominated
by the broad 9-13 $\mu$m emission; see Fig. 1) by a factor of $\sim$0.87. The
final continuum-subtracted images have been boxcar smoothed with a 2x2 window.
Figures 6 and 7 show the resulting PAH2$-$Si3, SiC$-$Si3, and Q1$-$Q4
continuum-subtracted images as well as the PAH2$-$SiC subtracted image, which
were flux-calibrated using the synthetic photometry (case ii) and the standard
stars (case i), respectively.

The continuum-subtracted images obtained using the standard stars for flux calibration
(case i; Fig. 7) seem to be of poorer quality than those obtained using the synthetic
photometry (case ii; Fig. 6). This is due to the use of standard stars observed on
different dates (at slightly different weather conditions), which produces non-systematic
flux uncertainties that vary ramdonly among the GTC/\-Ca\-na\-ri\-Cam filters; the
relative flux differences between GTC/\-Ca\-na\-ri\-Cam and {\it ISO} vary between
$\sim$10 and 30\% (depending on the filter; see Table 1). This is clearly evidenced by
the resulting PAH2$-$Si3 and SiC$-$Si3 images (case i; Fig. 7) that show a strong excess
emission closer to the central star. This excess of emission in the inner regions of the
nebula is an artifact due to a bad removal of the dust continuum emission at 9.8 $\mu$m
(Si3 filter; see Fig. 3). The resulting PAH2$-$Si3 and SiC$-$Si3 images obtained with the
case ii) flux calibration (Fig. 6) do not display such excess emission, indicating that
the Si3 dust continuum contribution is subtracted correctly. Taking into account this and
the non-IR variability of IC 418 (Fig. 2) in the following we thus make use of the
continuum-subtracted images obtained using the synthetic photometry (case ii) flux
calibration). 

\begin{figure*}
%Figure 6
\centering
\includegraphics[angle=0,scale=.58]{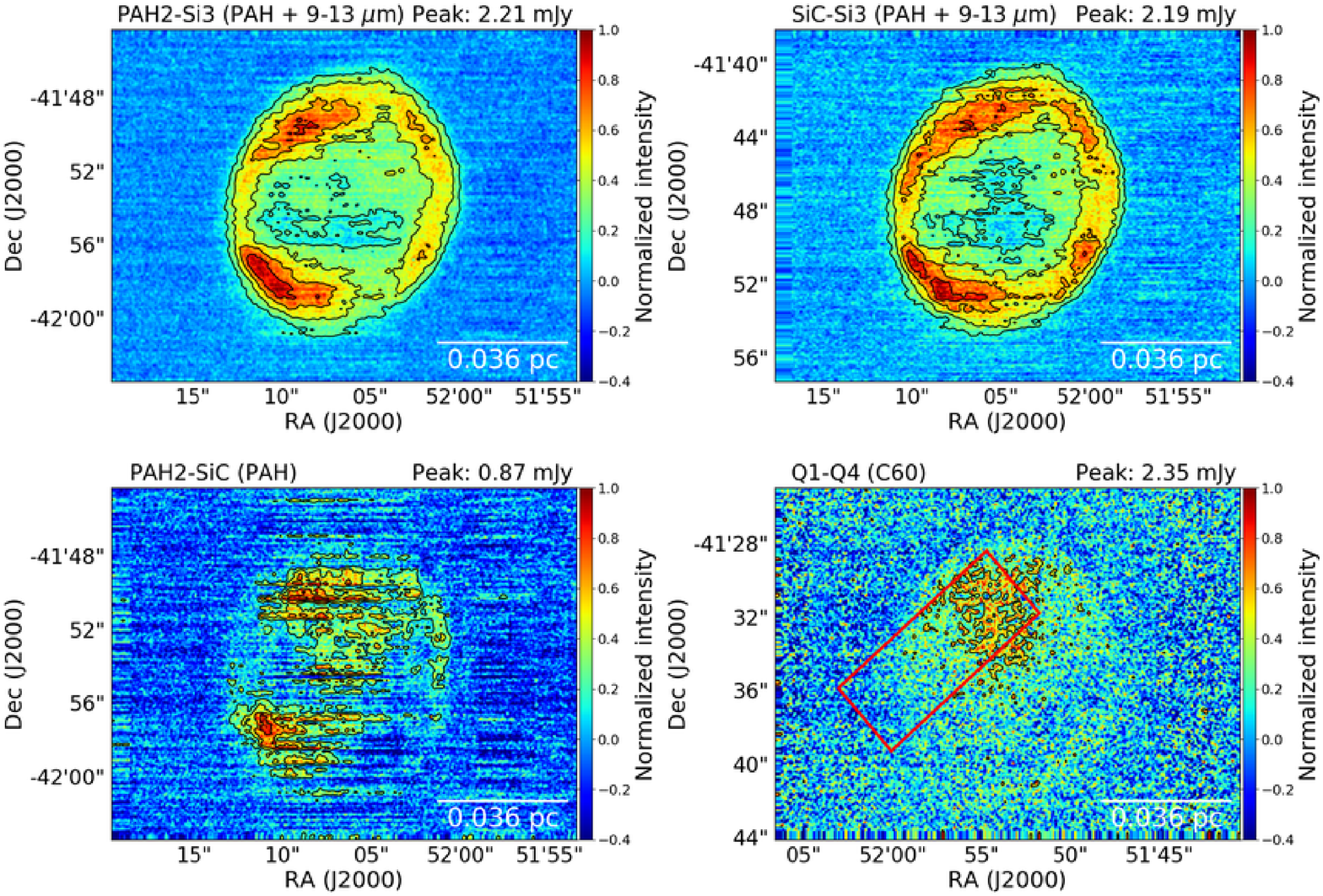}
\caption{Contour maps of the mid-IR GTC/\-Ca\-na\-ri\-Cam PAH2$-$Si3, SiC$-$Si3,
PAH2$-$SiC, and Q1$-$Q4 images of IC 418. The flux calibration was performed
using the synthetic photometry (case ii in the text) and all contour maps have
been normalized to the peak flux in the image. In the PAH2$-$Si3, SiC$-$Si3, and
PAH2$-$SiC images, contours range from 0.2 to 0.8 with 3 steps of 0.2 each. In
the Q1$-$Q4 image only one contour at 0.4 is displayed. Note that the position
and the size of {\it Spitzer}/IRS SH slit is indicated by the red box
(dimension: 4.7" x 11.3") superimposed to the Q1$-$Q4 image. \label{fig6}}
\end{figure*}

\begin{figure*}
%Figure 7
\centering
\includegraphics[angle=0,scale=.58]{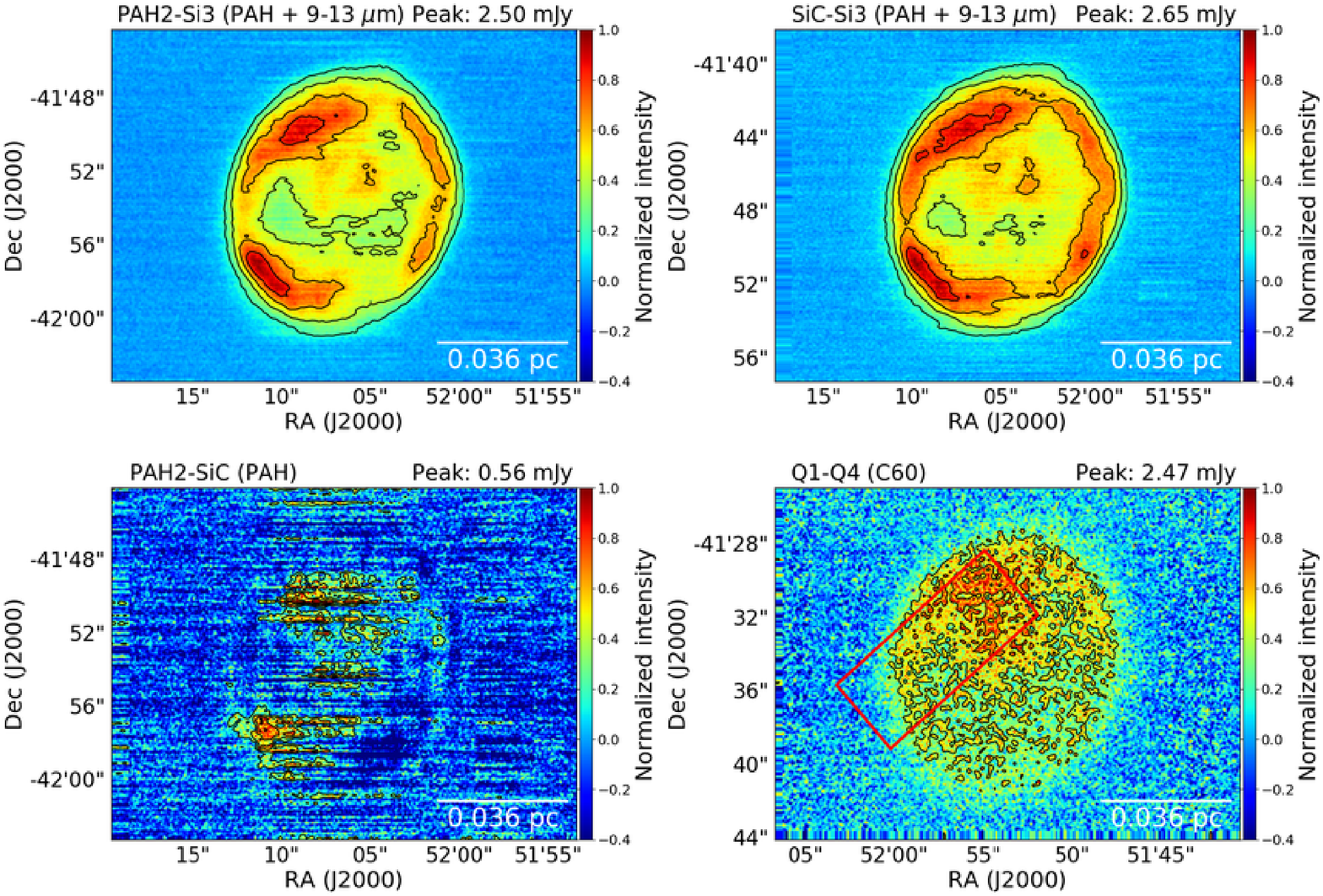}
\caption{Contour maps of the mid-IR GTC/\-Ca\-na\-ri\-Cam PAH2$-$Si3, SiC$-$Si3,
PAH2$-$SiC, and Q1$-$Q4 images of IC 418 when calibrating with the standard
stars (case i in the text). The notation is the same as in Figure 6 but the
contours are slightly different. In particular,
only two contours at 0.4 and 0.6 are displayed in the Q1$-$Q4 image. \label{fig7}}
\end{figure*}

\section{Results} \label{sec:resu}

\subsection{Fullerene emission} \label{subsubsec:fulle}

The Q1 and Q4 images (see Fig. 3) show a similar ring-like structure, which indicates
that fullerenes may be co-spatial with the dust continuum emission (e.g., they could be
attached to the dust grains) or that both images are dominated by the dust continuum
emission. IC 418 displays a very low C$_{60}$-to-continuum ratio (e.g.,
F$_{17.4}$/F$_{cont}$ $\sim$ 0.06, see below) and the similar spatial distribution seen
in the Q1 and Q4 images already suggests that both images are likely dominated by the
strong dust continuum emission. The continuum-subtracted fullerene emission (the Q1$-$Q4
image in Fig. 6) is very weak (at $\sim$4$\sigma$ from the mean background level) and
mainly located at the northeast, extending from the inner (even near the central star) to
the outer regions of the nebula. Due to the low C$_{60}$-to-continuum ratio we cannot
discard additional weaker fullerene emission from rest of the ring-like structure seen in
the Q1 image. Indeed, changing by $\sim$5-20$\%$ the factor used to subtract the Q4
continuum contribution, one can recover the ring-like structure in the resulting Q1$-$Q4
image. Alternatively, the emission may originate in an ellipsoid around the central star
and we see such distribution due to projection effects (see e.g., Manchado 2004).

We note that the {\it Spitzer}/IRS SH slit only covers part of the nebula, while the ISO
SWS diaphragm covers the entire nebula (see Pottasch et al. 2004 and Otsuka et al. 2014
for more details). In Figure 6, we display the Q1$-$Q4 continuum-subtracted image of IC
418 overlaid with the {\it Spitzer}/IRS SH slit. Interestingly, the {\it Spitzer} slit
just covers the northeast region of the nebula, where the residual fullerene emission
seems to be located. This could explain why {\it Spitzer} clearly detected the fullerene
IR features in IC 418 and it would suggest that the residual fullerene emission observed
could be real. We have measured the integrated fluxes of the C$_{60}$ bands at 17.4 and
18.9 $\mu$m (with SMART) in the {\it Spitzer} and {\it ISO} spectra (see Fig. 5). In
Table 2, we list the central wavelengths and integrated fluxes for these two C$_{60}$
fullerene bands in both {\it Spitzer} and {\it ISO} spectra. The integrated fluxes of
both C$_{60}$ bands in the {\it ISO} spectrum are four times higher than those of the
{\it Spitzer} one, as expected because of the larger {\it ISO} aperture and the fact that
weaker fullerene emission could be uniformly distributed along the nebula (Figs. 6 and
7). We measure C$_{60}$ F(17.4)/F(18.9) flux ratios of 0.50$\pm$0.12 and 0.39$\pm$0.24 in
the {\it Spitzer} and {\it ISO} spectra, respectively\footnote{We note that Otsuka et al.
(2014) measured a C$_{60}$ F(17.4)/F(18.9) flux ratio of 0.46$\pm$0.03 in the IC 418 {\it
Spitzer} spectrum, which agrees well, within the errors, with the one derived here.};
both ratios are similar within the errors. The feature-to-continuum ratios of the
C$_{60}$ 17.4 and 18.9 $\mu$m bands, however, are slightly higher in the {\it Spitzer}
spectrum (0.063$\pm$0.004 and 0.113$\pm$0.007, respectively) than in the {\it ISO}
(0.037$\pm$0.008 and 0.089$\pm$0.009, respectively) spectrum, which again might indicate
that the excess of fullerene 17.4 $\mu$m emission is real. Indeed, the integrated fluxes
measured in the Q1$-$Q4 and SiC$-$Si3 images, along the {\it Spitzer} SH slit, are
$\sim$2.09$\times$10$^{-18}$ and 3.07$\times$10$^{-17}$ Wcm$^{-2}$, which are $\sim$3.5
and 1.2 times higher than the C$_{60}$ 17.4 $\mu$m and 9$-$13 $\mu$m integrated fluxes
measured in the {\it Spitzer} spectrum (Table 2). Otherwise, we cannot discard the
possibility that the intensity of fullerene emission at 17.4 $\mu$m has varied spatially.

%\floattable
\begin{deluxetable}{lccccccccccc}
\tablenum{2}
\tablecaption{Mid-IR C$_{60}$ features in the PN IC 418\tablenotemark{a}} 
\tablehead{
\multicolumn{2}{c}{{\it Spitzer} spectrum} & \multicolumn{2}{c}{{\it ISO} spectrum} & \\
\hline
\colhead{$\lambda$} & \colhead{Flux} & \colhead{$\lambda$} & \colhead{Flux} \\
\colhead{($\mu$m)} & \colhead{(Wcm$^{-2}$)} & \colhead{($\mu$m)} & \colhead{(Wcm$^{-2}$)} }
\startdata
17.35 & 6.05e-19($\pm$0.72) & 17.32 & 1.31e-18($\pm$0.52) \\
18.97 & 1.20e-18($\pm$0.15) & 18.97 & 3.35e-18($\pm$0.75) \\
\enddata
\tablenotetext{a}{Estimated flux errors (between brackets) are always less than $\sim$30$\%$-40$\%$ (estimated by SMART).}
\label{tab:features}
\end{deluxetable}

The synthetic photometry through the GTC/\-Ca\-na\-ri\-Cam filters PAH2, SiC, and Q1 from
the {\it Spitzer} spectrum gives values of 11.42, 13.17, and 29.55 Jy, respectively,
while by summing the flux in our GTC/\-Ca\-na\-ri\-Cam images, in these three filters
along the {\it Spitzer} aperture (see Fig. 6), we get integrated fluxes of 15.90, 17.56,
and 31.61 Jy, respectively. The fluxes measured in our GTC/\-Ca\-na\-ri\-Cam images in
the filters covering the C$_{60}$ 17.4 $\mu$m (Q1) and the broad 9-13 $\mu$m (SiC)
emissions are thus about 7\% and 25\% higher than those from the {\it Spitzer} spectrum;
28\% higher for the narrower PAH2 filter.

Even if the flux calibration errors in our GTC/\-Ca\-na\-ri\-Cam data are difficult to
estimate, this also suggests a 17.4 $\mu$m excess emission in our GTC/\-Ca\-na\-ri\-Cam
observations compared with the {\it ISO} ones, which is in line with the apparent 17.4
$\mu$m excess emission in the {\it Spitzer} spectrum. 

Finally, it is to be noted here that we have assumed that the 17.4 $\mu$m
flux observed in IC 418 is mostly due to C$_{60}$ fullerenes, being not
contaminated by PAH-like 17.4 $\mu$m emission. This is because PAH-like features
in C$_{60}$-PNe are very weak (see e.g., Garc\'{\i}a-Hern\'andez et al. 2012;
Otsuka et al. 2014), unlike the case of RNe and RCB stars whose mid-IR spectra
are PAH-like dominated. The PAH-like 17.4$\mu$m band is correlated with the
PAH-like 16.4$\mu$m band (Bern\'e \& Tielens 2012) and the latter is only
present in RNe and RCBs (with PAH-like dominated spectra), being completely
absent in all C$_{60}$-PNe with different degrees of PAH-like emission; from
almost no PAH-like (e.g., Tc 1) to relatively stronger PAH-like emission (e.g.,
K 3-62). Among C60-PNe, the normalized line profiles of the 17.4 and 18.9 $\mu$m
C$_{60}$ bands are very similar and the  F(17.4$\mu$m)/F(18.9$\mu$m) ratio
remains almost constant (e.g., Otsuka et al. 2014). This already suggests that
possible PAH-like 17.4$\mu$m is very likely a minor contribution to the observed
17.4$\mu$m flux. On the other hand, with a slight extrapolation of Boersma et
al. (2010)'s Figure 5 (showing a correlation between the  6.2/11.3$\mu$m and
16.4/11.3$\mu$m PAH-like flux ratios), the 16.4$\mu$m band is predicted to have
an intensity $\sim$5\% that of IC 418's 11.3$\mu$m band (i.e.,
F(16.4$\mu$)$\sim$1.37$\times$10$^{-15}$ Wm$^{-2}$), which is consistent with
the feature's absence (i.e., well below our estimated upper limits). Other
PAH-like features in this spectral region are thus expected to be weaker. For
example, in the outer regions of the RNe NGC 7023, the 17.4$\mu$m band is 100\%
due to PAH-like species and F(17.4$\mu$m)/F(16.4$\mu$m)=0.35 (Bern\'e \& Tielens
2012), while the latter ratio is typically ~$\sim$0.05 in no C$_{60}$-PNe (e.g.,
Boersma et al. 2010). In the range F(17.4$\mu$m)/F(16.4$\mu$m)$\sim$0.35$-$0-05,
the expected PAH-like 17.4um contribution to the 17.4$\mu$m flux observed in IC
418, would be $\sim$27$-$191 times lower. Thus, assuming that the shape of the
spectrum remains constant over the nebula (this is at least true for the ISO and
Spitzer spectra), we conclude that the observed 17.4$\mu$m is very likely
uncontaminated by PAH-like blends.

\subsection{11.3 $\mu$m PAH-like + 9-13 $\mu$m carrier emission} \label{subsubsec:carr}

The Si3 image displays the strongest emission near the IC 418 central star and weaker
emission (mainly at the northeast and southeast of the nebula) from the outer ring-like
structure. The PAH2 and SiC images, however, show the strongest emission from the outer
ring-like structure with the weaker emission near the central star, where the Si3
emission is maximum (see Figs. 3 and 4). The 11.3 $\mu$m PAH-like $+$ 9-13 $\mu$m carrier
emission is obtained by doing the PAH2$-$Si3 and SiC$-$Si3 subtractions (see Fig. 6). The
PAH2$-$Si3 and SiC$-$Si3 continuum-subtracted images display an identical spatial
distribution. The similar morphology together with the very weak PAH-like 11.3 $\mu$m feature
(see Fig. 1) suggest that the emission is dominated by the 9$-$13 $\mu$m carrier in both
filters. The latter is confirmed by the PAH2-SiC subtraction (see below). The 9$-$13
$\mu$m carrier emission is distributed around the outer ring-like structure, with the
strongest emission (at more than 13$\sigma$ from the mean background level in Fig. 6) at
the northeast (and southeast) sides of the nebula. 

The spatial distribution of the 9-13 $\mu$m carrier emission is very similar to the dust
continuum emission at 20.5 $\mu$m (Q4), which suggests that the 9$-$13 $\mu$m carrier is
co-spatial with the dust grains emitting at 20.5 $\mu$m. On the other hand, by comparing
the Si3 and SiC images with the IC 418 Hubble Space Telescope images in several optical
nebular emission lines (see Fig. 8), we find that the dust continuum emission at 9.8
$\mu$m is very similar to the [O\,{\sc iii}] nebular emission at 502 nm, indicating that
dust grains coexist with ionized material. We also find that the 9-13 $\mu$m carrier $+$
11.3 $\mu$m PAH-like emission is somehow similar to the [N\,{\sc ii}] nebular emission at 658
nm, indicating that these molecules may coexist with ionized material in the outer
regions of the nebula.

\begin{figure*}
%Figure 8
\centering
\includegraphics*[angle=0,scale=.58]{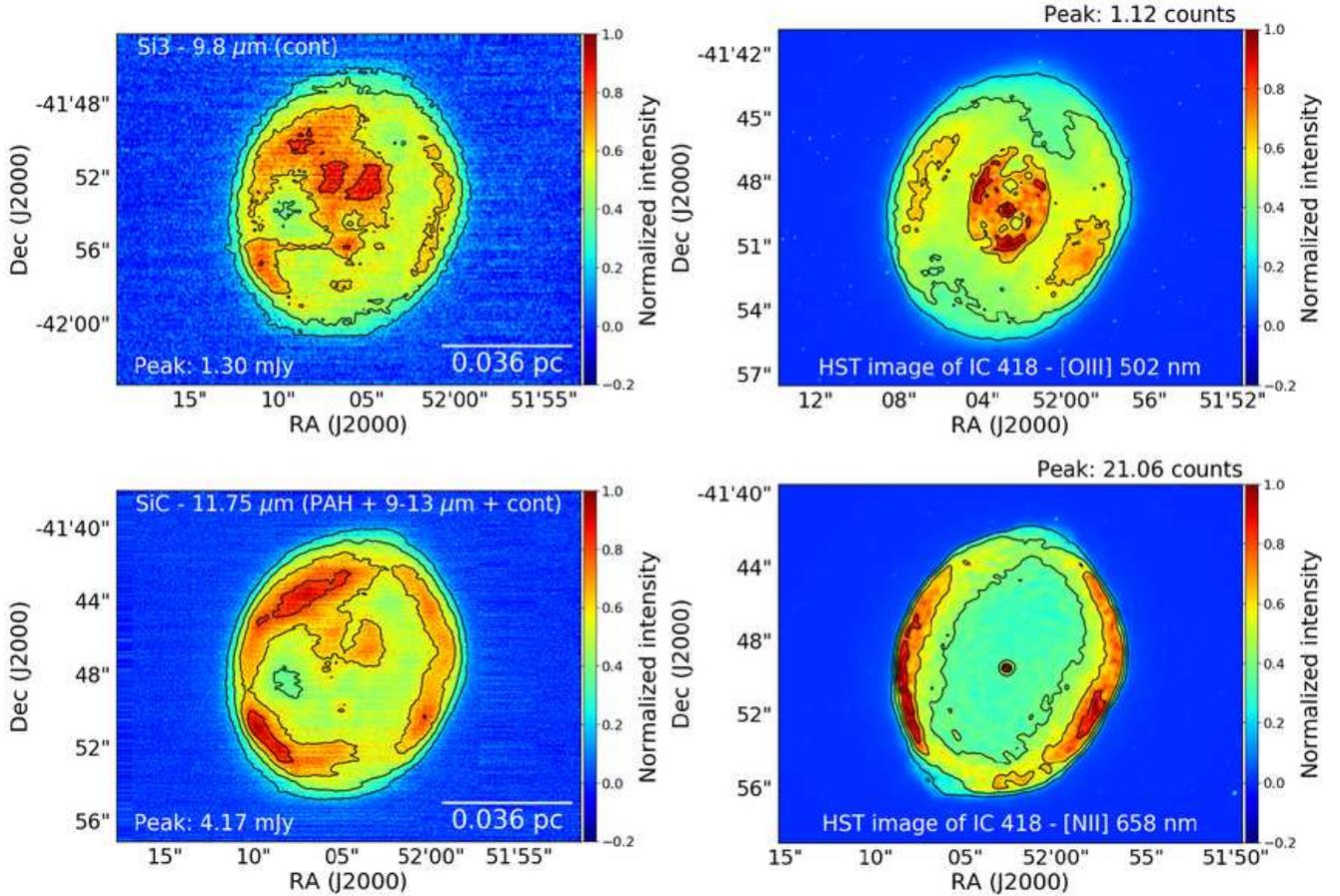}
\caption{Contour maps of the C$_{60}$-PN IC 418 in the mid-IR GTC/\-Ca\-na\-ri\-Cam
filters at 9.8 and 11.75 $\mu$m (left panels) and the optical HST F502N and F658N filters
at 502 and 658 nm (right panels), respectively. All contour maps have been normalized to the
peak flux in the image and contours range from 0.2 to 0.8 with 3 steps of 0.2
each. North is up, east is left. Note the similarity between the continuum
emission at 9.8 $\mu$m and [O\,{\sc iii}] nebular emission at 502 nm, and also
between the emission at 11.75 $\mu$m (9-13 $\mu$m carrier
$+$ 11.3 $\mu$m PAH-like emission $+$ continuum) and [N\,{\sc ii}] nebular
emission at 658 nm. \label{fig8}}
\end{figure*}

Interestingly, we could separate the PAH-like emission at 11.3 $\mu$m from the PAH2 image by
doing the PAH2$-$SiC subtraction (see Fig. 6). The PAH2$-$SiC image displays a less
extended residual emission with intensity peaks (at $\sim$5$\sigma$ from the mean
background level in Fig. 6) at the northeast and southeast sides of the nebula. This
suggests that the ring-like structure seen in the PAH2$-$Si3 and SiC$-$Si3
continuum-subtracted images is mostly due to the 9$-$13 $\mu$m carrier emission and that
the aromatic-like component coexist with the 9$-$13 $\mu$m carrier.

\subsection{Intensity profiles} \label{subsubsec:inten}

Normalized intensity profiles through the individual images (Si3, PAH2, SiC, Q1, and Q4
filters) and subtracted images (PAH2$-$Si3, SiC$-$Si3, PAH2$-$SiC, and Q1$-$Q4) at
position angles (P.A.) of 70$^\circ$ and 110$^\circ$ from east in a clockwise direction
are shown in Fig. 9. These P.A. were selected to trace the residual fullerene emission
(P.A.=70$^\circ$) as well as the ring-like structure along the major nebular axis
(P.A.=110$^\circ$). The profiles are made by doing a cross-cut through the x-axis and a
block average of 50 lines through the y-axis. We have chosen an arbitrary number of lines
that corresponds to 4 arcsec.

\begin{figure*}[t]
%Figure 9
\centering
\includegraphics[angle=0,scale=.63]{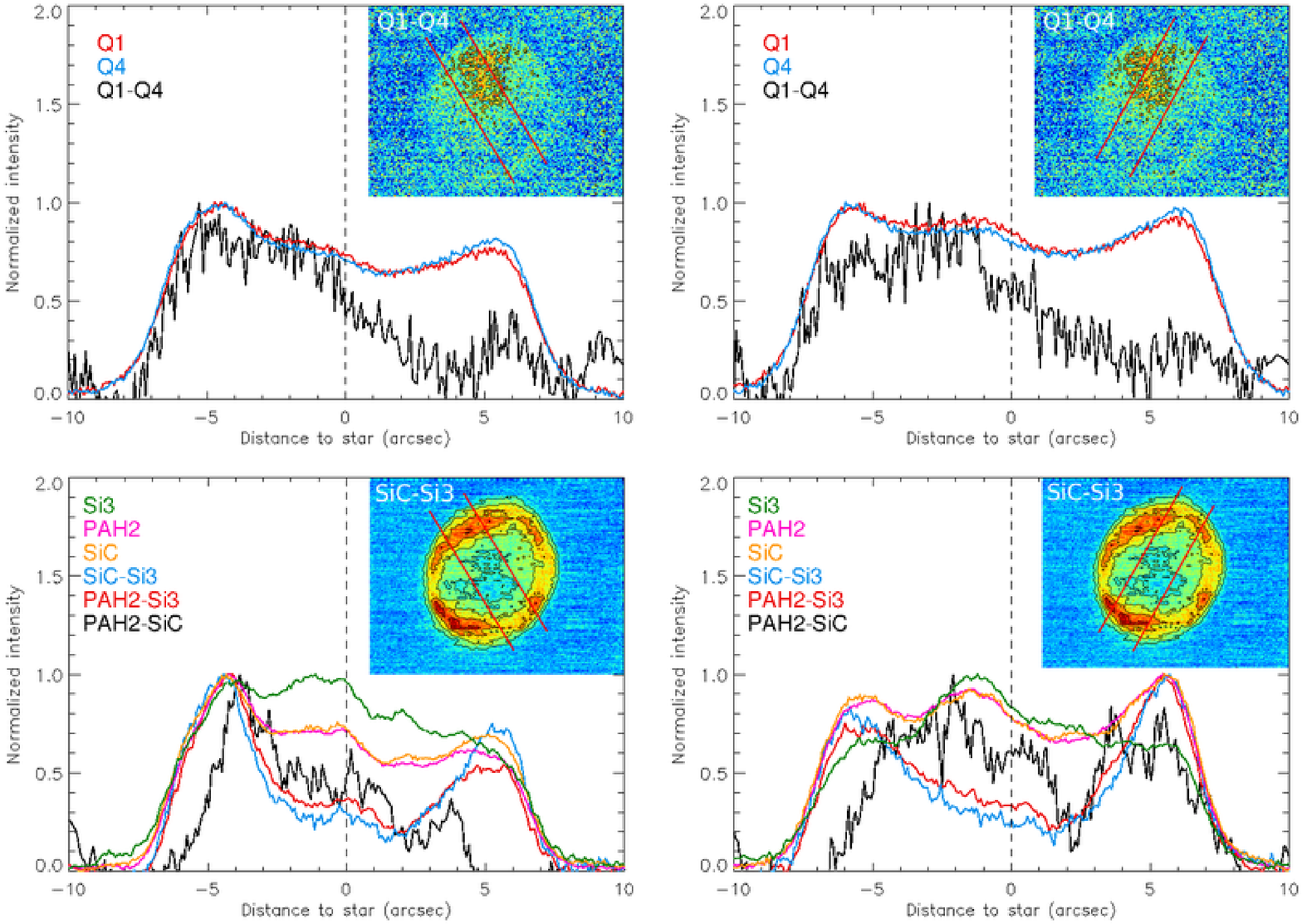}%
\caption{Normalized intensity profiles along the IC 418 nebula at P.A. of
70$^\circ$ (left panels) and 110$^\circ$ (right panels) from east as a function
of the distance to the central star. The P.A. and the 4 arcsec width are
indicated in the inset panels. The top panels show the intensity profiles in the
Q1 (C$_{60}$ $+$ continuum at 17.65 $\mu$m; in red), Q4 (continuum at 20.5
$\mu$m; in blue), and continuum-subtracted Q1$-$Q4 (C$_{60}$; in black) images.
The lower panels show the intensity profiles in the Si3 (continuum at 9.8
$\mu$m; in strong green), SiC and PAH2 (both containing 9-13 $\mu$m carrier $+$
11.3 $\mu$m PAH-like emission $+$ continuum; in brown and pink, respectively),
PAH2$-$SiC (11.3 $\mu$m PAH-like emission; in black), and continuum-subtracted
SiC$-$Si3 and PAH2$-$Si3 (9-13 $\mu$m carrier $+$ 11.3 $\mu$m PAH-like emission;
in blue and red, respectively) images, which were calibrated using the synthetic
photometry. The vertical dashed line marks the center of the nebula (or central star
position). \label{fig9}}
\end{figure*}

The intensity profiles along the Q1 and Q4 images have the same shape (independently of
the P.A.) because the continuum contribution is likely dominating the emission in both
filters, while the intensity profile along the Q1$-$Q4 show the location and extension of
the residual fullerene emission (upper panels in Fig. 9). The same behaviour is seen for
the PAH2 and SiC filters and their intensity profiles are very similar (they mainly trace
the outer ring-like structure) because both filters just cover the same features but with
different filter width. However, both PAH2 and SiC intensity profiles are different to
the Si3 intensity profile (covering the continuum at 9.8 $\mu$m), which shows the maximum
intensities at $\sim$1.2" from the central star ($\sim$1512 AU at 1.26 kpc; Morisset \&
Georgiev 2009; see Fig. 9). It is notable that the outer ring-like structure is clearly
seen in the SiC$-$Si3 and PAH2$-$Si3 profiles, where the continuum contribution has been
greatly subtracted from the SiC and PAH2 profiles. Finally, the PAH2$-$SiC profiles show
the residual excess emission towards the northeast of the nebula (P.A.=70$^\circ$) as
well as the even weaker (and more homogenous) emission along the major nebular axis
(P.A.=110$^\circ$).

The residual fullerene emission (see Fig. 9) mainly extends (almost uniformly) from the
central star to $\sim$7" at the northeast side of the nebula ($\sim$8820 AU), peaking at
$\sim$4.5-5.5" ($\sim$5670-6930 AU); a secondary much weaker (and tentative) peak is also
present at the opposite side of the nebula at $\sim$5.5-6" ($\sim$6930-7560 AU). The 9-13
$\mu$m carrier (+ weak 11.3 $\mu$m PAH-like) emission is distributed along the ring-like
structure with emission peaks at similar distances from the central star position (at
$\sim$4.5" and $\sim$5.5" at the northeast and southwest sides of the nebula) with
extremely weak emission in the inner regions. The weak 11.3 $\mu$m PAH-like emission, on the
contrary, seems to be located in a more internal region than the 9-13 $\mu$m carrier,
peaking at $\sim$4" ($\sim$5040 AU) at the northeast side of the nebula (see the
bottom-left panel in Fig. 9); another peak emission of similar strength (which is not
traced by the intensitiy profiles at P.A.=110$^\circ$ displayed in the bottom-right panel
in Fig. 9) is seen at a similar distance towards the southeast.

\section{Discussion} \label{sec:disc}

The obtained mid-IR images provide information about the spatial distribution of
the fullerene and PAH-like features as well as of the 9-13 $\mu$m feature carrier in
a fullerene-rich PN such as IC 418. 

Fullerene emission at 17.4 $\mu$m extends from the central star to $\sim$8820 AU
at the northeast side of the nebula (the maximum intensity is at $\sim$6300 AU).
This is comparable to the case of the other known fullerene-rich extended PN Tc
1, where lower spatial resolution {\it Spitzer} observations suggest that the
8.5 $\mu$m fullerene emission peaks at 6400-9700 AU from the central star
(Bernard-Salas et al. 2012)\footnote{A complete ring-like structure of the
C$_{60}$ 8.5$\mu$m emission, at a similar distance of $\sim$8000 AU, seems to be
confirmed by preliminary (non-continuum subtracted) higher spatial resolution
images of Tc 1 from Gemini-S/Trecs (Golriz et al., in prep.).}\nopagebreak.
Furthermore, Bernard-Salas et al. (2012) found that fullerene emission at 8.5
$\mu$m and PAH-like emission at 11.3 $\mu$m seem to be displaced from the central
star and peak at opposite directions, while the IR emission at 9 $\mu$m (dust
continuum) and 12 $\mu$m (the 9-13 $\mu$m carrier) are less extended and
centered on the Tc 1's central star. In the case of PN IC 418, however, it seems
that fullerene 17.4 $\mu$m and PAH-like 11.3 $\mu$m are not necessarily co-spatial
but both emissions have peaks towards the northeast of the nebula; although at
sligthly different distances from the central star (at $\sim$6300 and $\sim$5040
AU, respectively). However, the emission may originate in an ellipsoid around
the central star and we see such distribution due to projection effects (see
e.g., Manchado 2004). In IC 418, the dust continuum emission at 9.8 $\mu$m is
also extended and peaking near the central star (at $\sim$1512 AU), but the 9-13
$\mu$m carrier is mainly distributed in the outer ring-like structure. Thus, it
is not actually clear if the 12 $\mu$m emission in Tc 1 might be dominated by
the dust continuum or by the 9-13 $\mu$m carrier. In our mid-IR images of IC
418, it seems clear that the 9-13 $\mu$m carrier emission dominates over the
adjacent dust continuum emission.

The interpretation of our IC 418 GTC/\-Ca\-na\-ri\-Cam mid-IR images regarding
the dominant fullerene formation process in PNe is somewhat contradictory
(not conclusive). The presence of fullerene emission near the central star,
where the 9-13 $\mu$m emission is much weaker, would suggest that C$_{60}$ may
be a photo-product of the 9-13 $\mu$m carrier. This would agree with the
idea of fullerenes being formed from the photochemical processing of HAC-like
materials (e.g., Garc\'{\i}a-Hern\'andez et al. 2010), assuming that the 9-13
$\mu$m carrier is related to HAC-like dust grais. However, fullerenes seem to
coexist with the 9-13 $\mu$m carrier at the northeast outer regions of the
nebula; something that could indicate that fullerenes may be attached to dust
grains or well mixed with the 9-13 $\mu$m carrier. An alternative interpretation
may be that fullerenes are not directly linked to the 9-13 $\mu$m carrier.
Finally, another exciting interpretation (perhaps more consistent with the
fullerene emission observed) is that other fullerene-based species are
contributing to the observed 17.4 $\mu$m emission (see also Subsection
4.1); with ionized and/or more resistant fullerene-based species located in the
inner regions near the central star and neutral (and/or less resistant)
fullerenes distributed in the outer ring-like structure. More complex
fullerene-based species can emit at this wavelength. For example, laboratory IR
spectra of fullerene-based species like fullerene-PAHs adducts display spectral
features strikingly similar to those from C$_{60}$ (and C$_{70}$) fullerenes
(e.g., Garc\'{\i}a-Hern\'andez et al. 2013, 2016; Cataldo et al. 2014, 2015).
Furthermore, a recent theoretical study of the vibrational modes of hydrogenated
fullerenes (fulleranes; C$_{60}$H$_{m}$, m=2-36) shows that the four mid-IR
bands (at $\sim$7.0, 8.5, 17.4, and 18.9 $\mu$m) of the C$_{60}$-skeletal
vibrations are stronger in fullerenes at low hydrogenation (m=2-8) and the
astronomically observed IR bands assigned to C$_{60}$ could be due to
fullerenes/fulleranes mixtures (Zhang et al. 2017). Interestingly, most of
C$_{60}$H$_{m}$ (m=2-8) display stronger emission features at 17.4 $\mu$m
(i.e., higher 17.4 $\mu$m/18.9 $\mu$m band ratios) than C$_{60}$ and fulleranes
with very low H-content could potentially explain the intriguing spatial
distribution of the residual 17.4 $\mu$m emission seen in IC 418. Indeed,
the presence of fulleranes with very low H-content would be consistent with the
non-detection of $\sim$3.4$-$3.6 $\mu$m emission bands in the fullerene-rich PNe
Tc 1 and M 1-20 (D\'{\i}az-Luis et al. 2016); the strength of these fullerane
bands depends on the number of active C-H bonds, being much stronger in the
highly hydrogenated ones (Sadjadi 2017; priv. comm.; see also Zhang et al.
2017).

We note that our mid-IR images do not provide definitive proof about the
dominant excitation mechanism of the fullerene emission (e.g., thermal or
fluorescence emission); as it is also found in the literature (e.g.,
Garc\'{\i}a-Hern\'andez et al. 2012; Bernard-Salas et al. 2012; Otsuka et al.
2014; Brieva et al. 2016). Our observations may suggest that fullerenes could be
attached to dust grains due to the considerable similarity between the spatial
distribution of the IR emission in the Q1 and Q4 filters. In thermal excitation,
fullerenes should closely follow the dust distribution and show a similar
behaviour with respect to the distance to the central star. However, the
temparature of thermally excited fullerene molecules in PNe range from $\sim$200
to 600 K (e.g., Garc\'{\i}a-Hern\'andez et al. 2012) and it would be of only 33
K in IC 418 (Equation (3) in Garc\'{\i}a-Hern\'andez, Rao \& Lambert 2012).
Thus, the spatial distribution of fullerenes in IC 418 (as well as in Tc 1)
could favor fluorescence excitation. Fluorescence models, on the contrary,
cannot fully reproduce the emission of the fullerene IR bands in PNe and
contamination of the C$_{60}$ band strengths from UIR or other species like
fullerene-clusters has been suggested (e.g., Bernard-Salas et al. 2012, 2013;
Brieva et al. 2016). This would be consistent with the idea of other
fullerene-based species like fulleranes with very low H-content emitting at 17.4
$\mu$m, as tentatively suggested by the IC 418 mid-IR observations presented
here (see above).

In short, our GTC/\-Ca\-na\-ri\-Cam observations of IC 418 thus show that higher
sensitivity mid-IR images as well as spatially resolved mid-IR spectra along the nebula
(e.g., using the upcoming James Webb Space Telescope, JWST) are needed in order to
confirm the spatial distribution of the fullerene-, aromatic-, and aliphatic-like species
and to learn about fullerene formation and excitation in the complex circumstellar
environments around PNe. 

\section{Summary/Conclusions} \label{sec:summ}

We have presented for the first time high spatial resolution
(seeing-limited) narrow-band mid-IR images of a C$_{60}$-containing PN (IC
418); covering the fullerene band at 17.4 $\mu$m, the PAH-like feature at
11.3 $\mu$m, and the broad 9-13 $\mu$m aliphatic-like feature as well as their
adjacent continua. We have studied the relative spatial distribution of
these complex species with the main aim of getting some observational
constraints on the formation process of fullerenes in the H-rich circumstellar
environments around PNe.

A similar ring-like extended structure is seen in all narrow-band filters,
with the exception of the dust continuum emission at 9.8 $\mu$m, which peaks
much closer to the central star. The continuum-subtracted images show a clear
ring-like extended emission structure for the broad 9-13 $\mu$m feature carrier,
while only weak residual emission (at $\sim$4$-$5$\sigma$ from the mean
background level) is recovered for the PAH-like and fullerene emissions at 11.3 and
17.4 $\mu$m, respectively. The 11.3$\mu$m PAH-like emission seems to partially follow
the ring-like structure seen at other wavelengths, while the 17.4 $\mu$m
C$_{60}$-like emission is mainly located at the northeast, extending from the
central star to the outer regions of the nebula.

The residual C$_{60}$ emission, if real, might have several interpretations
(not conclusive) regarding the dominant fullerene formation process in PNe.
We speculate that other more complex fullerene compounds are likely
contributing to the observed 17.4 $\mu$m emission. In this context, hydrogenated
fullerenes with very low H-content seem to be potential candidates for the
possible contaminants, which otherwise could explain the present inconsistency
between the fullerene IR band ratios observed in PNe and the predictions of
thermal/fluorescence excitation models as well as the lack of
$\sim$3.4$-$3.6 $\mu$m emission features in fullerene-rich PNe.

We conclude that higher sensitivity mid-IR images and spatially resolved
spectroscopic observations (e.g., by using the upcoming JWST) are necessary to
confirm the spatial distribution of the fullerene-, aromatic-, and
aliphatic-like species in fullerene-rich PNe. The JWST is thus the ideal
astronomical facility to understand the formation and excitation of fullerenes
in circumstellar environments.

\acknowledgments

We acknowledge Omaira Gonz\'alez Mart\'in for her help during the data reduction with RedCan. JJDL,
DAGH, and AMT also acknowledge support provided by the Spanish Ministry of
Economy and Competitiveness (MINECO) under grant AYA$-$2014$-$58082$-$P. DAGH
was also funded by the Ram\'on y Cajal fellowship number RYC$-$2013$-$14182. EV
acknowledges support from the Spanish Ministry of Economy and Competitiveness
(MINECO) under grant AYA$-$2014$-$55840P. This work has been partially supported
by MINECO grant AYA2016-78994-P.

\software{SMART (http://isc.astro.cornell.edu/smart). IRAF (Tody 1986, Tody 1993)}

\end{document}